\documentclass[12pt]{iopart}
\usepackage{amssymb}

\begin{document}

\title{Analytical model for laser-assisted recombination of hydrogenic atoms}

\author{Gavriil Shchedrin$ ^{1}$ and Alexander Volberg$ ^{2}$}

\address{$^{1}$Department of Physics and Astronomy and National Superconducting Cyclotron Laboratory,
Michigan State University, East Lansing, Michigan 48824, USA}
\address{$^{2}$Department of Mathematics, Michigan State
University, East Lansing, Michigan 48824, USA}
\ead{$^{1}$gavriil@msu.edu}
\ead{$^{2}$volberg@math.msu.edu}

\begin{abstract}
We introduce a new method that allows one to obtain an analytical cross section for the laser-assisted electron-ion collision in a closed form. As an example we perform a calculation for the hydrogen laser-assisted recombination. The $S$-matrix element for the process is constructed from an exact electron Coulomb-Volkov wave function and an approximate laser modified hydrogen state. An explicit expression for the field-enhancement coefficient of the process is expressed in terms of the dimensionless parameter $\kappa=
\left|{e\varepsilon_{0}}/{q\omega_{0}}\right|^{2}$, where $e$ and $q$ are the electron charge and momentum respectively, and $\varepsilon_{0}$ and $\omega_{0}$ are the amplitude and frequency of the laser field respectively. The simplified version of the cross section of the process is derived and analyzed within a soft photon approximation.
\end{abstract}

\maketitle

\section{Introduction}
Atomic scattering processes in the presence of a harmonic laser field is a significant part
of contemporary atomic physics \cite{col1}. A simple and accurate theoretical model for
laser-assisted atomic scattering processes would allow one to explain their distinctive
properties. The standard theoretical approach for laser-assisted electron-ion collisions consists of the construction of the $S$-matrix element for the corresponding process. The electron wave function in a combined Coulomb-laser field is given by the well-known Coulomb-Volkov state. The dressed state of the atom is described by the classical time-dependent perturbation series. Within these approximations the $S$-matrix element was derived and numerically analyzed.

The goal of this paper is the construction of a simple analytical model for an electron-ion collision
in a harmonic laser field. Namely, we have developed a new method that allows to derive an analytical expression for the cross section of a laser-assisted atomic scattering process in closed form. The new step consists in the using of the Bessel generating function as an argument for the Plancherel theorem. This allows one to perform the summation over the number of field harmonics so that the analytical expression for the cross section of the process can be explicitly written. As an example, we perform a calculation for a laser assisted hydrogen recombination process. An exact expression for the field-enhancement coefficient is given in terms of the dimensionless parameter $\kappa=\left|{e\varepsilon_{0}}/{q\omega_{0}}\right|^{2}$.

The article is organized as follows. In Section II, the problem is formulated with the necessary knowledge concerning the standard field-free hydrogen recombination process. In Section III we give the $S$-matrix element for the hydrogen laser-assisted recombination process based on the electron Coulomb-Volkov wave function and the laser modified hydrogen state. In Section IV we perform a detailed analysis of the Coulomb-Volkov wave function. In Sections V and VI we obtain partial and general differential cross sections of the laser-assisted hydrogen recombination. Section VII is devoted to the analysis of the soft photon approximation, in which a simplified differential cross section of the process is derived and analyzed. Section VIII summarizes the results and advantages of the developed method.

\newpage

\section{Laser-assisted hydrogen recombination}

The goal of this  paper is to provide an analytical expression for a laser assisted hydrogen recombination process,
\begin{equation}\label{R1}
p+e+L\hbar{\omega_{0}}\longrightarrow{H+\hbar{\omega}}.
\end{equation}
The additional term $L\hbar{\omega_{0}}$ ($L$ is the number of harmonics) indicates the presence of a laser field,
\begin{equation}\label{lf1}
\vec{\varepsilon}=\vec{\varepsilon}_{0}\sin{\omega_{0}t},
\end{equation}
and points out the conservation of quasienergy. Here $\vec{\varepsilon}_{0}$ is the amplitude of the field and $\omega_{0}$ is the field frequency.

The final result for the analytical cross section of the process (\ref{R1}) can be directly applied
to the following laser-assisted recombination processes of\\
i. hydrogenic atom (for instance, $He^{+}$)
\begin{equation}\label{R3}
^{Z}_{A}X^{+}+e+L\hbar{\omega_{0}}\longrightarrow{_{A}^{Z}X+\hbar{\omega}},
\end{equation}
ii. antihydrogen atom
\begin{equation}\label{R4}
\bar{p}+{e}^{+}+L\hbar{\omega_{0}}\longrightarrow{\bar{H}+\hbar{\omega}},
\end{equation}
iii. positronium atom formation
\begin{equation}\label{R5}
e+{e}^{+}+L\hbar{\omega_{0}}\longrightarrow{Ps+\hbar{\omega}}.
\end{equation}

Creation of an antihydrogen atom \cite{hbar1} has become a driving force in contemporary atomic physics.
Of particular interest in this area is the accuracy the value of the fine structure constant
$\alpha=e^{2}/\hbar{c}$ which can be improved by comparing atomic spectra of hydrogen and antihydrogen atoms.
Exploration of these spectra provides a unique opportunity for studying basic symmetries
in physics, including the CPT invariance \cite{CPT}.

The main reaction leading to the formation of the antihydrogen atom in modern experiments \cite{hbar1, hbar2} is the three-body process which was a subject of recent theoretical calculations \cite{pohl1, rob1}. The laser assisted antihydrogen recombination process (\ref{R4}), being dominated by the three-body process, was studied experimentally \cite{gab1} and theoretically \cite{brn1, Li1}, but the analytical expression for this process has not been found prior to this work.

The differential cross section of the reaction for the
standard field-free hydrogen recombination process is well known due to the principle of detailed balance
between the differential cross section of photo-recombination, $d\sigma_{fi}/d\Omega_{f}$, and that of photo-ionization, $d\sigma_{if}/d\Omega_{i}$, \cite{landau1}:
\begin{equation}\label{p1}
\frac{d\sigma_{fi}}{d\Omega_{f}}=\frac{k^{2}}{q^{2}}\frac{d\sigma_{if}}{d\Omega_{i}},
\end{equation}
where $k$ and $q$ are the momenta of the outgoing photon and electron, respectively.

The well-known expression for the differential photo-ionization cross section $d\sigma_{if}/d\Omega_{i}$ is given by \cite{stob, AB}:

\begin{equation}\label{p2}
\frac{d\sigma_{if}}{d\Omega_{i}}=2^{4}\frac{e^{2}\omega}{m{\xi^{4}}}\left(\frac{\xi^{2}}{\xi^{2}+1}\right)^{5}
\frac{e^{-4\xi{arccot{\xi}}}}{1-e^{-2\pi\xi}}\frac{(1-\cos^{2}{\theta})}{(W_{0}-\hbar\omega)^{2}},
\end{equation}
where $W_{0}$ is the ionization potential of the hydrogen atom from the continuum threshold to the ground state. Here $\omega$ is the frequency of the photon emitted at the angle $\theta$ relative to the incoming electron, and $m$ and $e$ are the mass and electrical charge of the electron.
The dimensionless parameter $\xi$ is defined as
\begin{equation}\label{defxi}
    \xi={\frac{{Z\hbar}}{{a_{0}q}}}=\frac{\eta}{q},
\end{equation}
where $Z$ is the nuclear charge ($Z=1$), $a_{0}$ is the Bohr radius, and $\eta={Z\hbar}/a_{0}$

\section{The $S$-matrix element}

The $S$-matrix element describing the photo-recombination process
(\ref{R1}) in the presence of the laser field (\ref{lf1}) is given by \cite{AB, brn1, Li1}:
\begin{equation}\label{S1}
S=-ie\int{d{t}}{\langle{{\Psi}^{H}_{0}(\vec{r},t)|
(\vec{\epsilon}\cdot{\nabla})e^{-i(\vec{k}\cdot\vec{r}/\hbar-\omega{t})}
|\chi^{e}(\vec{r},t)\rangle}},
\end{equation}
where $\chi^{e}(\vec{r},t)$ is the Coulomb-Volkov wave
function of the electron in the field of the proton
and external laser field, $\vec{\epsilon}, \vec{k}$ and
$\omega$ are the polarization vector, momentum and frequency of the
emitted photon, respectively, and $\Psi^{H}_{0}(\vec{r},t)$ is the wave function of the hydrogen in the laser field $\vec{\varepsilon}$,  (\ref{lf1}).

Using the standard technique of transformation to the rotating frame we can obtain the electron wave function in the total Coulomb-laser field \cite{Tz1}
\begin{eqnarray}\label{pos1}
\chi^{e}(\vec{r},t)=e^{\frac{\pi{\xi}}{2}}\Gamma(1-i\xi)F(i\xi,1;i(qr-\vec{q}\cdot\vec{r})/\hbar)\\\times\nonumber
\exp\left[-\frac{ie}{\hbar{mc}}\int^{t}_{0}\vec{q}\cdot\vec{A}(\tau)d\tau-
\frac{ie^{2}}{2\hbar{mc^{2}}}\int^{t}_{0}A^{2}d\tau
+\frac{i\vec{q}\cdot\vec{r}}{\hbar}-
\frac{i{E_{i}}t}{\hbar}\right],
\end{eqnarray}
where $\vec{q}$ is the momentum of the incoming electron, $q=|\vec{q}|$, $r=|\vec{r}|$, $F(a,b;x)$ is the confluent hypergeometric function, $E_{i}$ is the initial kinetic energy of the incoming electron, and $c$ is the speed of light.

In the harmonic laser field (\ref{lf1}) the vector-potential $\vec{A}$ is defined as
\begin{equation}\label{vec1}
\vec{A}(t)=\frac{c\vec{\varepsilon}_{0}}{\omega_{0}}\cos{\omega_{0}t}\equiv{\vec{A}_{0}}\cos{\omega_{0}t}.
\end{equation}
In what follows we will use the velocity gauge (or $A\cdot{p}$ gauge). Therefore the term in the square brackets in
  (\ref{pos1}), $({ie^{2}}/{2\hbar{mc^{2}}})\int^{t}_{0}A^{2}d\tau$,
can be neglected \cite{Tz1}.
Evaluation of the remaining integral over $\tau$ in  (\ref{pos1}) yields
\begin{eqnarray}\label{pos2}
\chi^{e}(\vec{r},t)=\Gamma(1-i\xi)F(i\xi,1;i(qr-\vec{q}\cdot\vec{r})/\hbar)\\\nonumber
\times
\exp{i\left[\frac{\vec{q}\cdot\vec{r}}{\hbar}+
\frac{e\vec{q}\cdot\vec{\alpha}_{0}\sin{\omega_{0}t}}{m\hbar}-
\frac{E_{i}t}{\hbar}-
\frac{i\pi{\xi}}{2}\right]},
\end{eqnarray}
where $\vec{\alpha}_{0}=\vec{\varepsilon}_{0}/{\omega^{2}_{0}}$.

Using first-order perturbation theory for deriving the hydrogen wave function in the presence of the laser field we obtain \cite{brn1, Li1, landau1}

\begin{eqnarray}\label{ah1}
\Psi^{H}_{n}(\vec{r},t)=e^{-iW_{n}t/{\hbar}}\\\nonumber
\times\left\{{\psi_{n}(\vec{r})}
-\frac12\sum_{m\neq{n}}\left(
\frac{e^{i\omega_{0}{t}/\hbar}}{\omega_{mn}+\omega_{0}}+\frac{e^{-i\omega_{0}{t}/\hbar}}{\omega_{mn}-\omega_{0}}\right)
\langle{m|\frac{e\vec{A_{0}}\cdot\vec{p}}{\hbar{mc}\omega_{0}}|n\rangle}\psi_{m}(\vec{r})\right\},
\end{eqnarray}
where $\psi_{m}(\vec{r})$ is the wave function for the atomic electron in the field-free state $|m\rangle$ with energy $W_{m}$, and $\omega_{mn}=(W_{m}-W_{n})/\hbar$.
The summation in  (\ref{ah1}) is extended over the full set of atomic electron states in the absence of the laser field.
In the derivation of   (\ref{ah1}) it was assumed that none of the denominators were close to zero \cite{landau1}.

For the optical frequency of the laser we have $\omega_{0}\ll{\omega_{n0}}$ and therefore we get the following
expression for $\Psi^{H}_{0}(\vec{r},t)$, which holds true even over a  broader frequency range \cite{brn1}:
\begin{eqnarray}\label{ah2}
\Psi^{H}_{0}(\vec{r},t)=
e^{-iW_{0}t/\hbar}\left(1+\frac{ie(\vec{\varepsilon}_{0}\cdot\vec{r})}{\hbar{\omega_{0}}}\cos{\omega_{0}t}
\right)\psi^{H}_{0}(\vec{r}),
\end{eqnarray}
where \begin{equation}\label{C1}
\psi^{H}_{0}(\vec{r})=\sqrt{\frac{{Z^{3}}}{{\pi{a^{3}_{0}}}}}e^{-\eta{r}}\equiv{
C_{0}e^{-\eta{r}}}
\end{equation}

\section{The Coulomb-Volkov wave function}

In order to perform the time integration in the $S$-matrix element (\ref{S1}) we decompose the electron wave function $\chi^{e}(\vec{r},t)$ given by (\ref{pos2}) over the Bessel functions $J_{L}(z)$. For this purpose we introduce the Bessel generating function:
\begin{equation}\label{B1}
\exp(iz\sin{u})=\sum_{L=-\infty}^{L=+\infty}J_{L}(z)\exp(iLu).
\end{equation}
To simplify the $S$-matrix element (\ref{S1}) we use the recurrence relation for the Bessel functions:
 \begin{equation}\label{B2}
J_{L+1}(z)+J_{L-1}(z)=\frac{2L}{z}J_{L}(z).
\end{equation}
Performing the time integration in  (\ref{S1}) and with the aid of the Gauss theorem
we obtain the following expression for the $S$-matrix element in the dipole approximation ($(\vec{k}\cdot\vec{r})\ll{1}$)
\begin{equation}\label{S2}
S=-2\pi{i}\sum_{L=-\infty}^{L=+\infty}
f_{L}\delta(W_{0}+\hbar{\omega}-E_{i}+L\hbar{\omega_{0}}),
\end{equation}
where
\begin{equation}\label{f1}
f_{L}=e^{\frac{\pi{\xi}}{2}}\Gamma(1-i\xi){C_{0}}\left[J_{L}(z)\omega(L)\left(I_{1}+\frac{L}{z}I_{2}\right)\right].
\end{equation}
Here we have introduced the following notations:

\begin{equation}\label{z}
z=\frac{e(\vec{q}\cdot\vec{\alpha}_{0})}{{m\hbar}},
\end{equation}

\begin{equation}\label{I1}
I_{1}=\eta\hbar\int{d\vec{r}}{\psi}^{H}_{0}(\vec{r})\frac{(\vec{\epsilon}\cdot\vec{r})}{r}\chi^{e}(\vec{r}),
\end{equation}

\begin{equation}\label{I2}
I_{2}=\frac{ie}{\omega_{0}}\int{d\vec{r}}{{\psi}}^{H}_{0}(\vec{r})
\left(-(\vec{\varepsilon}_{0}\cdot{\vec{\epsilon}})+\eta\frac{(\vec{\varepsilon}_{0}\cdot\vec{r})
(\vec{\epsilon}\cdot\vec{r})}{r}\right)\chi^{e}(\vec{r}),
\end{equation}
and
\begin{equation}\label{om}
\hbar{\omega(L)}=E_{i}-W_{0}+L\hbar{\omega_{0}}\equiv{E_{i0}+L\hbar{\omega_{0}}}.
\end{equation}
Exploiting the well-known integral involving the confluent hypergeometric function \cite{AB, AM}
\begin{eqnarray}\label{I5}
\int{d\vec{r}}e^{i(\vec{q}-\vec{p})\cdot\vec{r}/\hbar-\eta{r}/\hbar}\frac{F(i\xi,1,i(qr-\vec{q}\cdot\vec{r}))}{r}
=4{\pi}\hbar^{2}\frac{[\vec{p}^{2}+(\eta-i{q})^{2}]^{-i{\xi}}}{[(\vec{q}-\vec{p})^{2}+\eta^{2}]^{1-i\xi}},
\end{eqnarray}
we get the following expression for  (\ref{I1}):
\begin{equation}\label{I3}
I_{1}=8\pi{i}\hbar^{4}(\vec{\epsilon}\cdot\vec{e}_{q})
\frac{\xi(1-i\xi)}{q^{2}(1+\xi^{2})^{2}}e^{-2\xi{arccot{\xi}}},
\end{equation}
and the corresponding expression for  (\ref{I2}):
\begin{eqnarray}\label{I4}
I_{2}=-8\pi{i}\hbar^{4}\frac{\xi{e}}{\omega_{0}}
\frac{e^{-2(1+\xi){arccot{\xi}}}}{q^3(1+\xi^{2})^2}
\left((\vec{\varepsilon}_{0}\cdot\vec{\epsilon})
-2(\vec{\epsilon}\cdot{\vec{e}_{q}})(\vec{\varepsilon}_{0}\cdot{\vec{e}_{q}})\frac{(2-i\xi)}{(1-i\xi)}\right),
\end{eqnarray}
where $\vec{e}_{q}\equiv{\vec{q}/q}$ is a unit vector.

\section{The partial cross section}

The partial cross section of the reaction (\ref{R1}) with fixed $L$ is given by
\begin{equation}\label{cs0}
d\sigma_{L}=2\pi\frac{e^{2}}{2q\omega(L)}
|f_{L}|^{2}\delta(W_{0}+\omega-E_{i}+L\hbar{\omega_{0}})
\frac{d^{3}k}{(2\pi)^{3}}.
\end{equation}
The integration over $\omega$ yields
\begin{eqnarray}\label{cs1}
\frac{d\sigma_{L}}{d\Omega}=
\frac{e^{2}C^{2}_{0}}{8{\pi}^{2}q(1-e^{-2\pi{\xi}})}
\left(E_{i}-W_{0}+L\hbar{\omega_{0}}\right)^{3}\\\nonumber
\times|J_{L}(z)|^{2}
\left(|I_{1}|^{2}+\frac{2Re(I_{1}I_{2})}{z}L+\frac{L^{2}}{z^{2}}|I_{2}|^{2}\right).
\end{eqnarray}
Here the constant $C_{0}$ was defined in (\ref{C1}).
The total cross section of the reaction (\ref{R1}) in the laser field (\ref{lf1}) is obtained with
\begin{equation}\label{cs2}
\frac{d\sigma}{d\Omega}=\sum^{L=+\infty}_{L=-\infty}\frac{d\sigma_{L}}{d\Omega}.
\end{equation}

\section{The summation procedure and cross section}

We introduce a new step that allows one to analytically sum up  the infinite series (\ref{cs2}). The summation over $L$ in (\ref{cs2}) is performed with the aid of the Plancherel theorem \cite{rudin},
\begin{equation}\label{pl1}
\frac{1}{2\pi}\int^{2\pi}_{0}{|f(x)|^{2}dx}=\sum^{n=+\infty}_{n=-\infty}|c_{n}|^{2},
\end{equation}
where $c_{n}$ are the Fourier coefficients of the function $f(x)$,
\begin{equation}\label{pl2}
c_{n}=\frac{1}{2\pi}\int^{2\pi}_{0}{f(x)e^{i\pi{nx}}}dx.
\end{equation}
The application of the Plancherel theorem to the Bessel generating function (\ref{B1}) leads to the following equalities \cite{bat}:
\begin{eqnarray}\label{B3}
\sum_{L=-\infty}^{L=+\infty}J^{2}_{L}(z)=1,
\end{eqnarray}
\begin{eqnarray}\label{B4}
\sum_{L=-\infty}^{L=+\infty}L^{2}J^{2}_{L}(z)=\frac{z^{2}}{2},
\end{eqnarray}
\begin{eqnarray}\label{B5}
\sum_{L=-\infty}^{L=+\infty}L^{2n-1}J^{2}_{L}(z)=0, n\in{{Z}_{+}}
\end{eqnarray}
\begin{eqnarray}\label{B6}
\sum_{L=-\infty}^{L=+\infty}L^{4}J^{2}_{L}(z)=\frac{z^{2}(4+3z^{2})}{8}.
\end{eqnarray}
Performing the summation over the photon polarizations
we obtain the closed analytical expression for the cross section of the process (\ref{R1}):
\begin{eqnarray}\label{cs3}
\frac{d\sigma}{d\Omega}=
\frac{e^{2}C_{0}^{2}E_{i0}^{3}}{8{\pi}^{2}q(1-e^{-2\pi{\xi}})}
\left\{|I_{1}|^{2}
+\left[\frac{|I_{2}|^{2}}{2}\right.\right.\\\nonumber
+\frac{3\hbar\omega_{0}z^{2}}{2E_{i0}^{2}}
\left(\frac{2E_{i0}\cdot{Re(I_{1}I_{2})}}{z}+
\hbar\omega_{0}|I_{1}|^{2}
\right)\\\nonumber
+\left.\left.\frac{\hbar^{2}\omega_{0}^{2}z^{2}(4+3z^{2})}{8E_{i0}^{3}}
\left(\frac{3E_{i0}|I_{2}|^{2}}{z^{2}}+\frac{2\hbar\omega_{0}
\cdot{Re(I_{1}I_{2})}}{z}\right)\right]\right\}.
\end{eqnarray}

Here $I_{1}$ specifies the space integral (\ref{I3}) for the field-free recombination process.
$I_{2}$ is the spatial integral (\ref{I4}) of the laser-assisted recombination process. The parameter
$z=e(\vec{q}\cdot\vec{\alpha}_{0})/{m\hbar}=e(\vec{q}\cdot{\vec{\varepsilon}_{0}})/{m\hbar{\omega^{2}_{0}}}$ was
introduced in (\ref{z}), where $\vec{q}$, $\varepsilon_{0}$, and $\omega_{0}$ denote the momentum of
the incoming electron, the amplitude of the laser field, and the field frequency, respectively.
The parameter  $E_{i0}\equiv{E_{i}-W_{0}}$ is the difference between the kinetic energy of the incoming electron and
the hydrogen ground state. The constant $C_{0}=\sqrt{{Z^{3}}/{\pi{a^{3}_{0}}}}$ represents the normalization constant (\ref{C1}) of the hydrogen wave function (\ref{ah2}) in the ground state.

It can be seen that in the zero-field limit, which corresponds to $z\equiv{0}$ and $I_{2}\equiv{0}$, the square brackets in (\ref{cs3}), which are responsible for the laser modified cross section, are identically zero.

To the best of our knowledge, such a closed expression for the laser assisted hydrogen recombination was not given in the literature up until now.

An explicit summation over the photon polarizations is given by the following
equalities
\begin{eqnarray}\label{pol1}
  \sum_{\lambda}\epsilon^{\lambda*}_{\mu}\epsilon^{\lambda}_{\nu}=\delta_{\mu\nu}, \\\nonumber
  \sum_{\lambda}(\vec{a}\cdot\vec{\epsilon}^{\lambda})^{2}=a^{2}(1-\cos^2{\vartheta}),
  \end{eqnarray}
where $\vartheta$ is the angle between the vector $\vec{a}$, and the momentum of an
outgoing photon $\vec{k}$.

The corresponding expressions for (\ref{I3}) and (\ref{I4})
with an explicit summation over the photon polarizations
are given by the following equalities
\begin{equation}\label{I5}
\sum_{\lambda}\left|I_{1}\right|^{2}=2^{6}\pi^{2}\hbar^{8}
\frac{\xi^{2}e^{-4\xi{arccot{\xi}}}}{q^{4}(1+\xi^{2})^{3}}(1-\cos^2{\theta}),
\end{equation}
\begin{eqnarray}\label{I6}
\sum_{\lambda}\left|I_{2}\right|^{2}=2^{6}\pi^{2}\hbar^{8}\frac{{e^{2}}\xi^{2}}{\omega_{0}^{2}}
\frac{e^{-4(1+\xi){arccot{\xi}}}}{q^{6}(1+\xi^{2})^{5}}\times\\\nonumber
\left[\varepsilon^{2}_{0}(1-\cos^2{\phi})(1+\xi^{2})
-4(\vec{\varepsilon}_{0}\cdot{\vec{e}_{q}})^{2}(2+\xi^{2})
+4(\vec{\varepsilon}_{0}\cdot{\vec{e}_{q}})^{2}(4+\xi^{2})(1-\cos^2{\theta})
\right],
\end{eqnarray}
and by the corresponding expression for the interference term
\begin{equation}\label{I7}
\sum_{\lambda}Re(I_{1}I_{2})=
2^{6}\pi^{2}\hbar^{8}\frac{{e}\xi^{2}}{\omega_{0}}
\frac{e^{-2(1+2\xi){arccot{\xi}}}}{q^{5}(1+\xi^{2})^{4}}
(\vec{\varepsilon}_{0}\cdot{\vec{e}_{q}})(4\cos^2{\theta}-3),
\end{equation}
where $\theta$ is the angle between the momentum of incoming electron $\vec{q}$ and the momentum of an outgoing photon $\vec{k}$. The angle $\phi$ is formed by vector $\vec{\varepsilon}_{0}$ and vector $\vec{k}$.

In order to give an estimation for the order of magnitude of the laser field correction to the recombination process we employ the Keldysh dimensionless tunneling parameter {\cite{kel}}:
\begin{equation}\label{est0}
\gamma=\frac{\omega_{0}}{e\varepsilon_{0}}\sqrt{2mW_{0}},
\end{equation}
which for characteristic numerical values $\varepsilon_{0}\equiv{|\vec{\varepsilon}_{0}}|\sim{5\times10^{7}}$ V/cm,
and
$\hbar{\omega_{0}}\sim{1}$ eV, and $W_{0}=13.6$ eV has the value $\gamma\sim{4}$.

The parameter $z$ (\ref{z}) can be easily expressed in terms of the Keldysh tunneling parameter and for the case $E_{i}\sim{W_{0}}=13.6$ eV has the following numerical value
\begin{equation}\label{z2}
    z=\frac{e(\vec{q}\cdot\vec{\alpha_{0}})}{{m\hbar}}=\frac{2\sqrt{E_{i}W_{0}}}{\gamma{\hbar{\omega_{0}}}}\sim{7},
\end{equation}
and the parameter $\xi$ given by (\ref{defxi}) can be estimated as
\begin{equation}\label{xi2}
    \xi={\frac{{Z\hbar}}{{a_{0}q}}}\sim{1}.
\end{equation}
Finally we get the following estimations for the field-influenced terms confined by the square brackets (\ref{cs3})
\begin{equation}\label{est1}
\frac{|I_{2}|^{2}}{|I_{1}|^{2}}
\sim{\kappa\equiv{
\left|\frac{e\varepsilon_{0}}{q\omega_{0}}\right|^{2}}=\frac{1}{\gamma^2}\frac{W_{0}}{E_{i}}\sim{\frac{1}{16}}},
\end{equation}
\begin{equation}\label{est2}
\frac{\hbar{\omega_{0}}z}{E_{i0}}
\frac{Re(I_{1}I_{2})}{|I_{1}|^{2}}
\sim{\kappa}
\sim{\frac{1}{16}},
\end{equation}
\begin{equation}\label{est3}
\frac{(\hbar{\omega_{0}}z)^2{|I_{1}|^{2}}}{E^2_{i0}{|I_{1}|^{2}}}
\sim{\kappa}
\sim{\frac{1}{16}},
\end{equation}
\begin{equation}\label{est4}
\frac{\hbar^{2}\omega_{0}^{2}(4+3z^{2})}{E_{i0}^{2}}
\frac{|I_{2}|^{2}}{|I_{1}|^{2}}
\sim{\kappa^2}
\sim{\frac{1}{256}},
\end{equation}
\begin{equation}\label{est5}
\frac{(\hbar\omega_{0})^3z(4+3z^{2})}{E_{i0}^{3}}
\frac{Re(I_{1}I_{2})}{|I_{1}|^{2}}
\sim{\kappa^2}
\sim{\frac{1}{256}}.
\end{equation}

The field-influenced terms, $\left|I_{1}\right|^{2}$ and $\left|I_{2}\right|^{2}$,
are positive and accompany the recombination process.
According to  (\ref{I7}) the interference term, $Re(I_{1}I_{2})$, is positive under the following conditions
\begin{eqnarray}
  (\vec{\varepsilon}_{0}\cdot{\vec{e}_{q}})>0, \theta\in\left(-\frac{\pi}{6};\frac{\pi}{6}\right)\bigcup
  \left(\frac{5\pi}{6};\frac{7\pi}{6}\right);\\\nonumber
  (\vec{\varepsilon}_{0}\cdot{\vec{e}_{q}})<0, \theta\in\left(\frac{\pi}{6};\frac{5\pi}{6}\right)\bigcup
  \left(-\frac{5\pi}{6};-\frac{\pi}{6}\right).
\end{eqnarray}
In the particular case with $(\vec{\varepsilon}_{0}\cdot{\vec{e}_{q}})\equiv{0}$ the interference terms are absent.

Based on the obtained estimations (\ref{est1})-(\ref{est5}) we do not see the predicted {\cite{Li1}} laser contraction of the hydrogen (antihydrogen) recombination process.

The basic estimations, (\ref{est1})-(\ref{est5}), show that the field influence terms are directly expressed in terms of the dimensionless parameter $\kappa=\left|{e\varepsilon_{0}}/{q\omega_{0}}\right|^{2}$. It is proportional to the intensity of the laser field, $|\varepsilon_{0}|^2$, and inversely proportional to the square of the field frequency, $1/\omega_{0}^2$. This is justified by the chosen electron (\ref{pos2}) and hydrogen (\ref{ah2}) laser modified states, where in both cases there is a dependence on the reciprocal field frequency.
The spatial integral of the laser-assisted recombination (\ref{I4}) is more sensitive to the momentum of the incoming electron than the field-free  spatial integral (\ref{I3}). This results in the inverse proportionality between $\kappa$ and  $q^{2}={2mE_{i}}$.

The parameter $\kappa\equiv{\left|{e\varepsilon_{0}}/{q\omega_{0}}\right|^{2}}={1}/{\gamma^2}{W_{0}}/{E_{i}}$ governs the
laser-assisted recombination process and points out a possible ways for an increasing a hydrogen laser-assisted recombination rate. An increase of the parameter $\kappa$  by lowering the parameter ${\gamma^2}$
is the most natural way to improve a recombination rate of a hydrogen atom in a laser field. Although it is
experimentally feasible, this path has to be examined carefully
due to an arising ionization processes of the produced hydrogen atom.
Ionization processes have not been considered in the proposed model and
therefore a quantitative treatment requires a full solution of the time-dependent Schr\"{o}dinger equation \cite{TDPT}.

The binding electron energy $W_{0}$ is a well-defined value and can not be used in the increasing the laser-assisted recombination rate. The ultimate possibility to increase the parameter $\kappa$ is decreasing the relative kinetic energy between electron and a proton $E_{i}$ for which one can achieve a reasonably small value which would mean improving a laser-assisted recombination rate. This fact has a clear physical explanation: for a quasi-static regime in the inverse to the laser-assisted recombination process, i.e. in the laser-assisted photo-ionization process, the energy of the emitted photo-electrons has a peak at zero value. In our case of laser-assisted recombination process this peak corresponds to a maxima value of photo-recombination rate.

\section{The soft photon approximation}

The cross section given by  (\ref{cs3}) is an exact result for the laser assisted hydrogen recombination process under the proposed approximations for the electron and hydrogen laser modified wave functions. To clarify the meaning of the obtained result we shall introduce a soft photon approximation. This allows one to reveal the meaning of each term in the expression for the cross section of the process given by (\ref{cs3}). With this approximation the frequency of the emitted photon is independent of the number of the field harmonics $L$:
\begin{equation}\label{om3}
\hbar\omega=E_{i}-W_{0}+L\hbar{\omega_{0}}\simeq{E_{i}-W_{0}}.
\end{equation}
This is justified as the low-frequency field limit, or the soft photon limit {\cite{brn1}}.
In this case we are able to produce a simple but non-trivial result.
In the soft photon approximation the corresponding expression for the partial "soft photon" cross section is given by
\begin{eqnarray}\label{cs4}
\frac{d\sigma^{sf}_{L}}{d\Omega}=
\frac{e^{2}C^{2}_{0}}{8{\pi}^{2}q(1-e^{-2\pi{\xi}})}
\left(E_{i}-W_{0}\right)^{3}\\\nonumber
\times|J_{L}(z)|^{2}
\left(|I_{1}|^{2}+\frac{2Re(I_{1}I_{2})}{z}L+\frac{L^{2}}{z^{2}}|I_{2}|^{2}\right).
\end{eqnarray}
Performing the summation (\ref{cs2}) over $L$ by means of the equalities (\ref{B3})-(\ref{B5}) we get the following simple result:
\begin{eqnarray}\label{cs5}
\frac{d\sigma^{sf}}{d\Omega}=
\frac{e^{2}C^{2}_{0}}{8{\pi}^{2}q(1-e^{-2\pi{\xi}})}
{E_{i0}^{3}}\left(|I_{1}|^{2}+\frac{|I_{2}|^{2}}{2}\right).
\end{eqnarray}
Here the first term $|I_{1}|^{2}$ corresponds to the standard laser-free recombination process whereas the second term $|I_{2}|^2$ is responsible for the field modified electron (\ref{pos2}) and hydrogen (\ref{ah2}) states. In the zero-field limit $I_{2}\equiv{0}$ and  (\ref{cs5}) recovers the standard field-free recombination cross section.
We have to note that within the limits of the present approximation (\ref{om3}) the interference terms $Re(I_{1}I_{2})$ are absent. Thus in this and only this particular case does  the square of the $S$-matrix element (\ref{S2}) equal the sum of the squares of its terms. So we have proved:

\newpage

{\bf Theorem}
{\it The S-matrix element}
$$
S=-2\pi{i}\sum_{L=-\infty}^{L=+\infty}
f_{L}\delta(W_{0}+\hbar{\omega}-E_{i}+L\hbar{\omega_{0}})
$$
{\it with}
$$
f_{L}=e^{\frac{\pi{\xi}}{2}}\Gamma(1-i\xi){C_{0}}\left[J_{L}(z)\omega(L)\left(I_{1}+\frac{L}{z}I_{2}\right)\right]
$$
{\it under the approximation}
$$
\hbar\omega=E_{i}-W_{0}+L\hbar{\omega_{0}}\simeq{E_{i}-W_{0}\neq{\hbar\omega(L)}},
$$
{\it equals to the sum of the squares of individual terms}
$$
|S|^{2}=\left|2\pi{e}^{\frac{\pi{\xi}}{2}}\Gamma(1-i\xi){C_{0}}\right|^{2}
\delta^2(W_{0}+\hbar\omega-E_{i})\omega^2\left(|I_{1}|^{2}+\frac{|I_{2}|^{2}}{2}\right).
$$

The Theorem reveals the meaning of each term in the expression for
 the laser-assisted hydrogen photo recombination given by (\ref{cs3}).
 The first term in the braces (\ref{cs3}), $|I_{1}|^{2}$, is responsible for the field-free recombination process.
 The first term in the square brackets (\ref{cs3}), $|I_{2}|^{2}/2$, reflects the laser-modified electron and hydrogen
 states and does not take into account the changes in the frequency of the emitted photon, according to the Theorem. The four terms in the parentheses (\ref{cs3}) are the only terms which contribute to the change of the frequency of the emitted photon in the process.

\section{Summary}

In this paper we have developed a new method that allows one to obtain an analytical cross section for the laser-assisted electron-ion collision. The standard $S$-matrix formalism is used for describing of the collision process. The $S$-matrix element is constructed from the electron Coulomb-Volkov wave function in the combined Coulomb-laser field, and the hydrogen laser modified state. By the aid of the Bessel generating function, the $S$-matrix element is decomposed into an infinite series of the field harmonics.

We have introduced a new step to obtain an analytical expression for the cross section of the process.
The main theoretical novelty is the application of the Plancherel theorem to the Bessel generating function. This allows one to obtain an analytical expression for the laser assisted hydrogen photo recombination process.

The laser-assisted hydrogen recombination process has been chosen in order to verify the proposed method. The field-enhancement coefficient is evaluated in an analytical way. The final expression for a laser-assisted hydrogen recombination process is presented by the sum of the field-free hydrogen cross section and the laser-assisted addition. The field-dependent terms are expressed through the dimensionless parameter $\kappa\equiv{
\left|{e\varepsilon_{0}}/{q\omega_{0}}\right|^{2}}$.

By introducing a soft photon approximation, based on the assumption of the independence between the frequency of the emitted photon and the field harmonics, the square of the $S$-matrix element is represented by the sum of the squares of its individual terms. We provide the proof of the corresponding theorem.

The developed method will allow one to reconsider a wide range of problems related to electron-ion collisions in an external field with the goal of obtaining an analytical expression for the cross sections of the corresponding process. The time-dependent problem generated by the infinite series of the Coulomb-Volkov wave function is exactly separated from the spatial dependence and thus can be analytically solved by the method.

One of the challenging problems in modern electron-ion collision theory is the calculation of a laser assisted three-body hydrogen formation process. As far as we know an analytical solution for this important process has not been found. The developed method allows one to obtain an analytical expression for this process.

\section*{Acknowledgements}
The authors are grateful to L. N. Labzowsky  and V. G. Zelevinsky for valuable discussions.
Also we are thankful to the Referees for a number of helpful suggestions for the improvement of this paper.

G.S. acknowledges the support from the Institute for the Quantum Sciences at Michigan State University.


\end{document}